\documentclass[%
 reprint,
 amsmath,amssymb,
 aps,
]{revtex4-2}

\usepackage{graphicx}
\usepackage{xcolor}
\usepackage{amsmath}
\usepackage{caption}
\usepackage{subcaption}
\usepackage{dcolumn}
\usepackage{bm}
\usepackage{caption}
\usepackage{subcaption}
\usepackage{hyphenat}
\usepackage{url}
\usepackage{svg}
\usepackage{lipsum}

\begin{document}

\title{Dynamics tuning with reservoir computer control}

\author{Krithikesh Ravishankar}
 \affiliation{Department of Applied Mathematics, University of Colorado at Boulder, Colorado 80309, USA}
 
 \author{Juan G. Restrepo}
\affiliation{Department of Applied Mathematics, University of Colorado at Boulder, Colorado 80309, USA}

 \author{Per Sebastian Skardal}
 \affiliation{Department of Mathematics, Trinity College, Hartford, Connecticut, 06106, USA}

\date{\today}


\begin{abstract}

We present an equation-free method for maintaining a dynamical system operating in a desired regime, even as system parameters are disturbed in such a way that qualitatively different dynamics would emerge. This method allows for real-time identification of changes made to a system's evolution equations induced by the parameter changes. If the equations governing the original system's dynamics are known, our method can allow one to determine the values of the parameters in real time. We illustrate our method with numerical examples using a reservoir computer control implementation. 
\end{abstract}

\maketitle

\section{Introduction} \label{sec:level1}

In many applications, it is desirable that a dynamical system operates in a regime associated with a prescribed set of parameters, such as a fixed point, a periodic orbit, or an attractor with specific statistical properties. Maintaining such dynamics in the face of unknown, potentially stochastic parameter variations and external disturbances can be challenging. In recent decades, a wide range of approaches have been developed to address such challenges, such as Ott-Grebogi-Yorke chaos control~\cite{ott1990controlling,shinbrot1993using}, robust control methods such as $H_{\infty}$~\cite{bacsar2008h} and linear–quadratic–Gaussian (LQG)~\cite{athans1971role, aastrom2012introduction} control, adaptive methods such as model reference adaptive control~\cite{nguyen2018model} and self-tuning regulators~\cite{aastrom1977theory}, stochastic formulations such as dual control~\cite{feldbaum1960dual}, and more recent techniques such as model predictive control~\cite{kouvaritakis2016model,schwenzer2021review} and reinforcement learning~\cite{sutton1998introduction,arulkumaran2017deep}. While each of these approaches offers important capabilities, they rely on assumptions that may limit their applicability in practice, such as the availability of a system model, linearity, or assumptions about the statistical properties of parameter fluctuations. These limitations continue to motivate the development of new algorithms and extensions of existing ones~\cite{chen2015disturbance, zhang2023optimal, yang2019suboptimal}. In this paper, we propose a data-driven method to maintain a system in a desired dynamical regime under both stochastic and deterministic parameter disturbances. Our approach requires that the response of the system to known perturbations can be observed, but does not require knowledge of the equations describing the evolution of the system or of the parameter fluctuations. When a model is known, our method allows for inference of the underlying parameter fluctuations in real time, even as the system is maintained in the desired regime.

In recent years, reservoir computing, originally introduced more than two decades ago~\cite{maass2002real,jaeger2004harnessing}, has become a useful tool for the analysis and forecasting of time-varying systems~\cite{lukovsevivcius2012reservoir,pathak2017using,pathak2018model,nakajima2021reservoir,cucchi2022hands}. Reservoir computers consist of a {\it fixed} high-dimensional dynamical system (the reservoir), a fixed input layer, and a trainable readout layer. After forcing the reservoir with an input signal, the weights of the readout layer are optimized to produce the desired signal. Since training is performed in a single least-squares minimization step and reservoir states provide short-term memory, reservoir computers offer some advantages over other machine learning architectures that require backpropagation~\cite{vlachas2020backpropagation}. Most uses of reservoir computing in dynamical systems are related to system prediction~\cite{pathak2018model} or attractor reconstruction~\cite{pathak2017using}, but their use in the design of control algorithms is increasingly attracting interest~\cite{canaday2021model,kent2024controlling,mandal2025adaptive, haluszczynski2023controlling}. In recent work~\cite{restrepo2024suppressing,skardal2023detecting}, we demonstrated how reservoir computers can be used to identify, in real time, external time-dependent disturbances applied to an unknown dynamical system, provided that the response of the system to known disturbances can be observed and used to train a reservoir computer. Here we consider a fundamentally different problem where parameter drift changes the governing vector field itself, potentially inducing bifurcations and transitions between dynamical regimes. We show that the reservoir framework can preserve target dynamics even under these structural dynamical changes while simultaneously inferring the drifting parameters in real time. Furthermore, we show how knowledge of these disturbances can be leveraged to apply a control signal to the system in order to suppress the effects of the parameter change.

This paper is structured as follows. In Sec.~\ref{sec2} we describe the problem and our solution based on reservoir computing. In Sec.~\ref{sec3} we illustrate our method with numerical simulations of various systems. Finally, in Sec.~\ref{sec4} we present our conclusions.

\section{Statement of the Problem} \label{sec2}

\subsection{Dynamics Tuning}

We consider a continuous-time dynamical system described by a state vector ${\bf x} \in \mathbb{R}^D$ that evolves according to the ordinary differential equation
\begin{align}
\frac{d {\bf x}}{dt} = {\bf F}({\bf x},{\bf p}_0),\label{original}
\end{align}
where ${\bf p}_0 \in \mathbb{R}^P$ is a  vector of $P$ parameters. Importantly, neither the right-hand side ${\bf F}$ nor the parameters ${\bf p}_0$ are assumed to be known. We assume that for the parameters ${\bf p}_0$ the system is in a stable attractor, and that it is desired to maintain the system in this attractor even if the parameters change. More precisely, suppose that the vector of parameters evolves as ${\bf p}(t)$, and the system is now given by 
\begin{align}
\frac{d {\bf x}}{dt} = {\bf F}({\bf x},{\bf p}(t)),\label{eq0}
\end{align}
which can be rewritten as 
\begin{align}
\frac{d {\bf x}}{dt} = {\bf F}({\bf x},{\bf p}_0) + {\bf g}({\bf x},t),\label{perturbed}
\end{align}
where ${\bf g}({\bf x},t)$ is the perturbation
\begin{align}
{\bf g}({\bf x},t) = {\bf F}({\bf x},{\bf p}(t)) - {\bf F}({\bf x},{\bf p}_0).
\end{align}
Emphasizing that no knowledge of the underlying dynamics ${\bf F}$ or the original parameters ${\bf p}_0$ is assumed, our goals are to (i) apply a control term to maintain the dynamics of the system as close as possible to the original dynamics (\ref{original}), and (ii) identify the perturbation ${\bf g}({\bf x},t)$.
We refer to this process as {\it dynamics tuning}, and we say that we tune the dynamics at parameters ${\bf p}_0$.

\subsection{Reservoir Computer Implementation}\label{implementation}

We present an implementation of dynamics tuning using reservoir computers based on our previous work in Refs.~\cite{restrepo2024suppressing,skardal2023detecting}. We assume that we can observe how the original system (\ref{original}) responds to {\it known} time-dependent perturbations during a training time interval $t \in [-T,0]$
\begin{align}
\frac{d \hat {\bf x}}{dt} = {\bf F}(\hat {\bf x},{\bf p}_0) + {\bf f}(t),\label{test}
\end{align}
where ${\bf f}$ is the {\it training function}. While the system is being perturbed with the training function, we collect a time series of the system state $\hat {\bf x}$, $\{ \hat {\bf x}(-T +j\Delta t) \}_{j = 0}^{T/ \Delta t}$ and the corresponding perturbations $\{ {\bf f}(-T +j\Delta t) \}_{j = 0}^{T/ \Delta t}$. These two time series are used to train a machine learning system to identify, at any given time, the training function given the system state $\hat{\bf x}$ and its recent history. In this paper we use reservoir computers, a machine learning framework with memory particularly useful for time-varying data~\cite{lukovsevivcius2012reservoir,pathak2017using,pathak2018model,nakajima2021reservoir,cucchi2022hands}. We implement the reservoir as a network of $M$ nonlinearly coupled units described by a state vector ${\bf r}^j$ for $j = 0, 1, \dots$. The reservoir states evolve as
\begin{align}
{\bf r}^{j+1} = \tanh(A{\bf r}^j + W_{\text{in}} \hat {\bf x}^j + {\bf 1}), \label{reservoir}  
\end{align}
where the $M\times M$ matrix $A$ encodes the internal structure of the reservoir, the $M\times D$ input matrix $W_{\text{in}}$ is used to force the reservoir with the state vector $\hat {\bf x}$, and ${\bf 1} = [1,1,\dots,1]^T \in \mathbb{R}^M$. After the reservoir is evolved according to Eq.~(\ref{reservoir}) for $j = 0, 1, 2, \dots, T/\Delta t$, the reservoir states $\{{\bf r}^j\}_{j=0}^{T/\Delta t}$ are collected, and a readout matrix $W_{\text{out}}$ is chosen so that ${\bf u}^j \equiv W_{\text{out}} {\bf r}^j\approx {\bf f}(-T+j\Delta t)$. More specifically, $W_{\text{out}}$ is chosen to minimize the cost function
\begin{align}
 \sum_{j=0}^{T/\Delta t} \| W_{\text{out}} {\bf r}^j - {\bf f}(-T+j \Delta t) \|^2 + \lambda \text{Tr}(W_{\text{out}}W_{\text{out}}^T),
\end{align}
where $\lambda$ is a regularization parameter. This least-squares optimization can be solved efficiently.

In our implementation, we use $M=200$ and construct the matrix $A$ by setting each entry to be nonzero with probability $6/M$; then, the value of each nonzero entry is selected uniformly and independently at random from $[-1,1]$; finally, the matrix is rescaled to ensure a spectral radius of $\rho=0.95$. The entries of the input matrix $W_{\text{in}}$ are chosen independently and uniformly at random from $[-0.01,0.01]$. Our regularization parameter is $\lambda=10^{-8}$.

Once the reservoir computer is trained to infer the applied perturbation from the system state and its recent history, the state vector ${\bf x}$ of the unknown system (\ref{perturbed})  can be used as input to the reservoir. The reservoir's output, ${\bf u} = W_{\text{out}} {\bf r}^j$, is taken as an approximation to the unknown disturbance ${\bf  g}$, ${\bf u} \approx {\bf g}$. In Refs.~\cite{restrepo2024suppressing,skardal2023detecting} we demonstrated how the reservoir is able to identify time-dependent forcing terms such as chaotic and stochastic signals. As we will see, even when trained with fixed time-dependent functions, the reservoir is able to identify perturbations which depend on the state of the system, ${\bf g}({\bf x},t)$, including those that result from parameter changes.

Once an approximation ${\bf u}$ to the perturbation ${\bf g}$ has been identified, it can be suppressed by the control scheme (see Ref.~\cite{restrepo2024suppressing} for more details, including a stability analysis)
\begin{align}
\frac{d{\bf x}}{dt} &= {\bf F}({\bf x},{\bf p}_0) + {\bf g}({\bf x},t) - \alpha {\bf v},\label{eq1}\\
\frac{d{\bf v}}{dt} &= \frac{1}{\tau}\left( {\bf u} - {\bf v}  \right),\label{eq2}
\end{align}
where $\alpha > 0$ is the control gain, ${\bf v}$ is a delayed version of ${\bf u}$, and $\tau$ controls the timescale of the delay. The motivation for this control scheme is that ${\bf u}$ can't be subtracted directly from Eq.~(\ref{perturbed}), since the reservoir would then identify ${\bf g} - {\bf u}$ as the perturbation, resulting in the output ${\bf u} \approx {\bf g} -  {\bf u}$, or ${\bf u} \approx {\bf g}/2$. Subtracting $\alpha {\bf u}$ suppresses the perturbation for large values of $\alpha$, but the control scheme becomes unstable. Using the delayed variable ${\bf v}$ increases the range of values of $\alpha$ over which the method remains stable. Since the reservoir output approximates the deviation from the original right-hand side of Eq.~(\ref{eq0}), which is ${\bf g} -\alpha {\bf v}$, the perturbation can be estimated in real time as
\begin{align}\label{identify}
{\bf g}({\bf x},t) \approx {\bf u}(t) + \alpha {\bf v}(t).
\end{align}

\begin{figure*}[t]
    \centering
    \includegraphics[width = 0.75\textwidth]{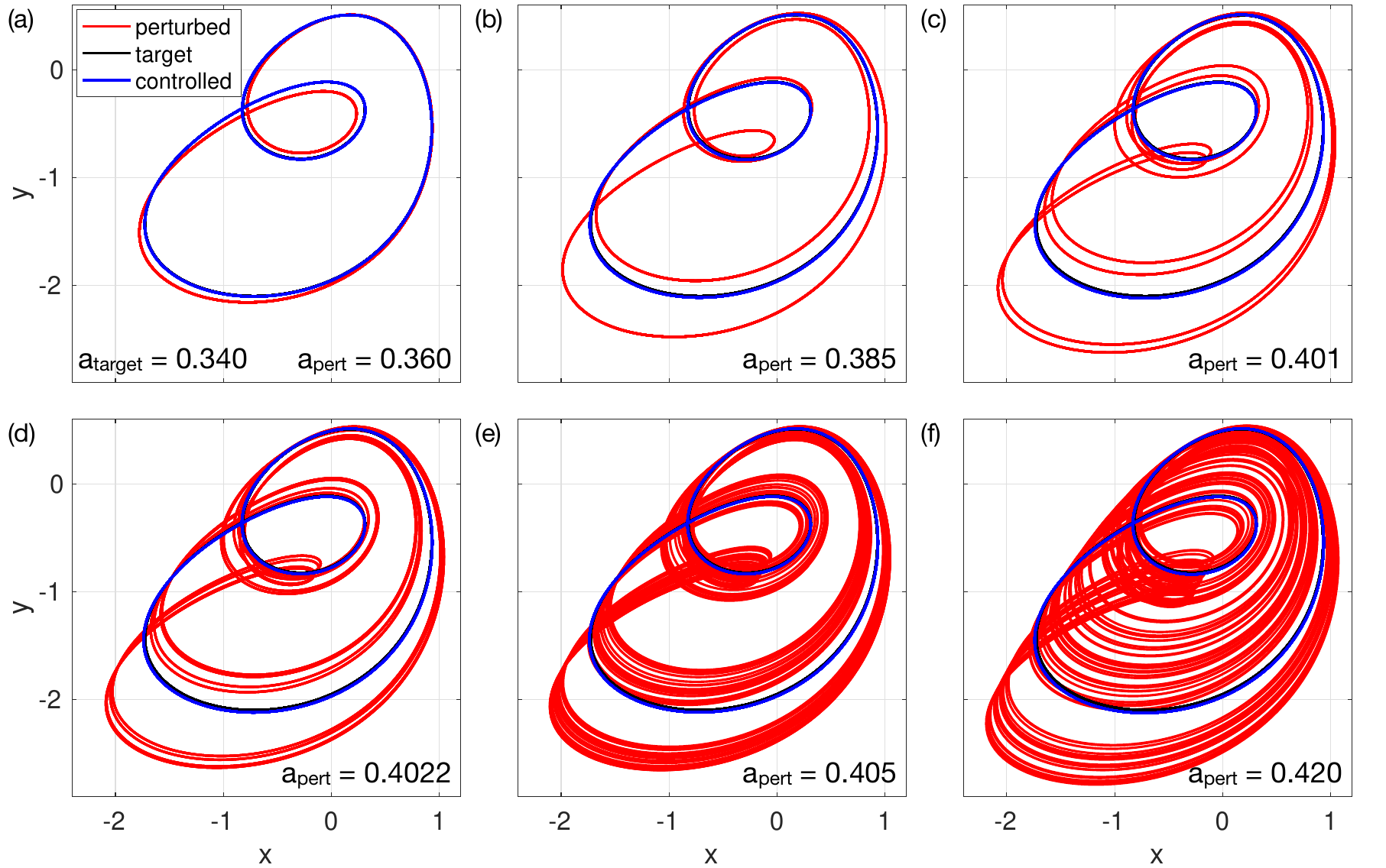}
    \caption{Attractor for the target system with $a = a_{\text{target}} = 0.34$ (black), for the system with $a = a_{\text{pert}}$ (red), and for the controlled system with $\alpha = 10$ and $\tau=1$ for $a_{\text{pert}} = 0.36$ (a), $0.385$ (b), $0.401$ (c), $0.4022$ (d), $0.405$ (e), and $0.420$ (f) (blue). Despite the perturbed attractor becoming chaotic, the controlled attractor is close to the periodic target attractor.}
    \label{fig:sprott}
\end{figure*}

So far, we have not required any knowledge of the function ${\bf F}$ or of the baseline parameters ${\bf p}_0$. If these are known, it is possible in some cases to determine the values of the parameters ${\bf p(t)}$ in real time. The perturbation is given by
\begin{align}
{\bf g}({\bf x}(t),t)&= {\bf F}({\bf x}(t),{\bf p}(t)) - {\bf F}({\bf x}(t),{\bf p}_0)\\& \approx {\bf u}(t) + \alpha {\bf v}(t),
\end{align}
and therefore, when 
\begin{align}\label{eq:G}
{\bf F}({\bf x}(t), {\bf p}) - {\bf F}({\bf x}(t),{\bf p}_0) \equiv {\bf G}({\bf x}(t),{\bf p})
\end{align}
is invertible as a function of ${\bf p}$, then 
\begin{align}
{\bf p}(t) \approx {\bf G}^{-1} \left({\bf x}(t), {\bf u}(t) + \alpha {\bf v}(t)\right).
\end{align}
With the real-time reservoir output ${\bf u}(t)$ and the associated function ${\bf v}(t)$, the parameters ${\bf p}(t)$ can be approximately determined from this equation. In the following, we will illustrate our method for dynamics tuning and parameter tracking with various examples. 

\section{Numerical Experiments} \label{sec3}

\subsection{Sprott-Linz Type F system}

We begin with the Sprott-Linz Type F system~\cite{sprott2000}, given by 
\begin{align}
\frac{dx}{dt} & = y+z,\\
\frac{dy}{dt} & = -x + a y,\\
\frac{dz}{dt} & = x^2 - z.
\end{align}
To illustrate our dynamics tuning algorithm, we first tune the dynamics at $a = a_{\text{target}} = 0.34$, a value that results in a periodic orbit [black curves in Figs.~\ref{fig:sprott}(a)-(f)]. For a time step we use $
\Delta t = 0.02$. Larger values of $a$ eventually result in chaotic behavior via a period-doubling bifurcation cascade [red curves in Figs.~\ref{fig:sprott}(a)-(f)]. We train the reservoir by using the training function 
$${\bf f}(t) = \left[\begin{array}{c}
0.1\cos((0.1+0.1r_x)t)\\
0.1\sin((0.1+0.1r_y)t)\\ 
0.1\cos((0.1+0.1r_z)t)
\end{array}
\right],
$$
where $r_x, r_y, r_z$ are three numbers chosen at random uniformly and independently from $[0,1]$. Then, for each value of $a = a_{\text{pert}}$ shown in Fig.~\ref{fig:sprott},  we implement the control scheme (\ref{eq1})-(\ref{eq2}).  Fig.~\ref{fig:sprott} shows the periodic orbit that exists for the base value $a_{\text{target}} = 0.34$ (black solid line), the attractor obtained from $a_{\text{pert}}$ (red curve), and the periodic orbit obtained using the dynamics tuning algorithm in Eqs.~(\ref{eq1})-(\ref{eq2}) with $\alpha = 10$ and $\tau = 1$ (blue curve). Even for values of $a = a_{\text{pert}}$ that produce chaotic dynamics, the controlled system remains very close to the target periodic orbit. To quantify the performance of this procedure, we define the {\it error} $H$ as the Hausdorff distance between the controlled trajectory $\mathcal{C}$ and the reference attractor $\mathcal{R}$ with $a = a_{\text{target}} = 0.34$, estimated numerically as
\begin{align}
H = \max\{ &\max_j\{\min_k||\hat {\bf x}(k\Delta t) - {\bf x}(j\Delta t)||\} ,\nonumber\\
&\max_k\{\min_j||{\bf x}(j\Delta t)-\hat {\bf x}(k\Delta t)||\} \}.\label{eq:meanerror}
\end{align}

\begin{figure}[t]
    \centering
    \includegraphics[width = 1.0\columnwidth]{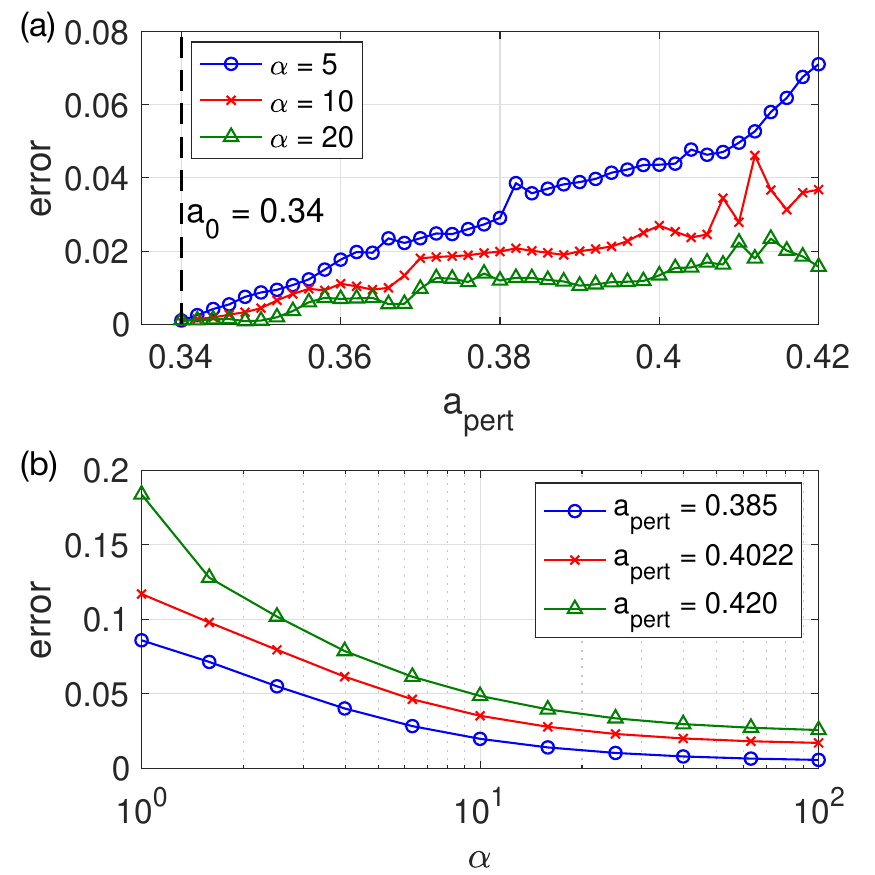}
    \caption{(a) Mean error [see Eq.~(\ref{eq:meanerror})] as a function of $a_{\text{pert}}$ for $\alpha = 5$ (blue circles), $\alpha = 10$ (red crosses), and $\alpha = 20$ (green triangles). The baseline parameter $a_0 = 0.34$ is indicated with a dashed vertical line. (b) Mean error as a function of $\alpha$ for $a_{\text{pert}} = 0.385$ (blue circles), $a_{\text{pert}} = 0.4022$ (red crosses), and $a_{\text{pert}} = 0.420$ (green triangles).}
    \label{fig2}
\end{figure}

In Fig.~\ref{fig2}(a) we plot the mean error as a function of $a_{\text{pert}}$ for $\alpha = 5, 10,$ and $20$. The baseline value $a_{\text{target}} = 0.34$ is indicated by a vertical dashed line. The error increases as $|a_{\text{pert}} - a_{\text{target}}|$ increases, but is reduced as the control gain parameter $\alpha$ is increased [cf. Eqs.~(\ref{eq1})-(\ref{eq2})]. Fig.~\ref{fig2}(b) shows how the mean error decreases as $\alpha$ is increased for three specific values of $a_{\text{pert}}$.

\begin{figure}[t]
    \centering
    \includegraphics[width = 1.0\columnwidth]{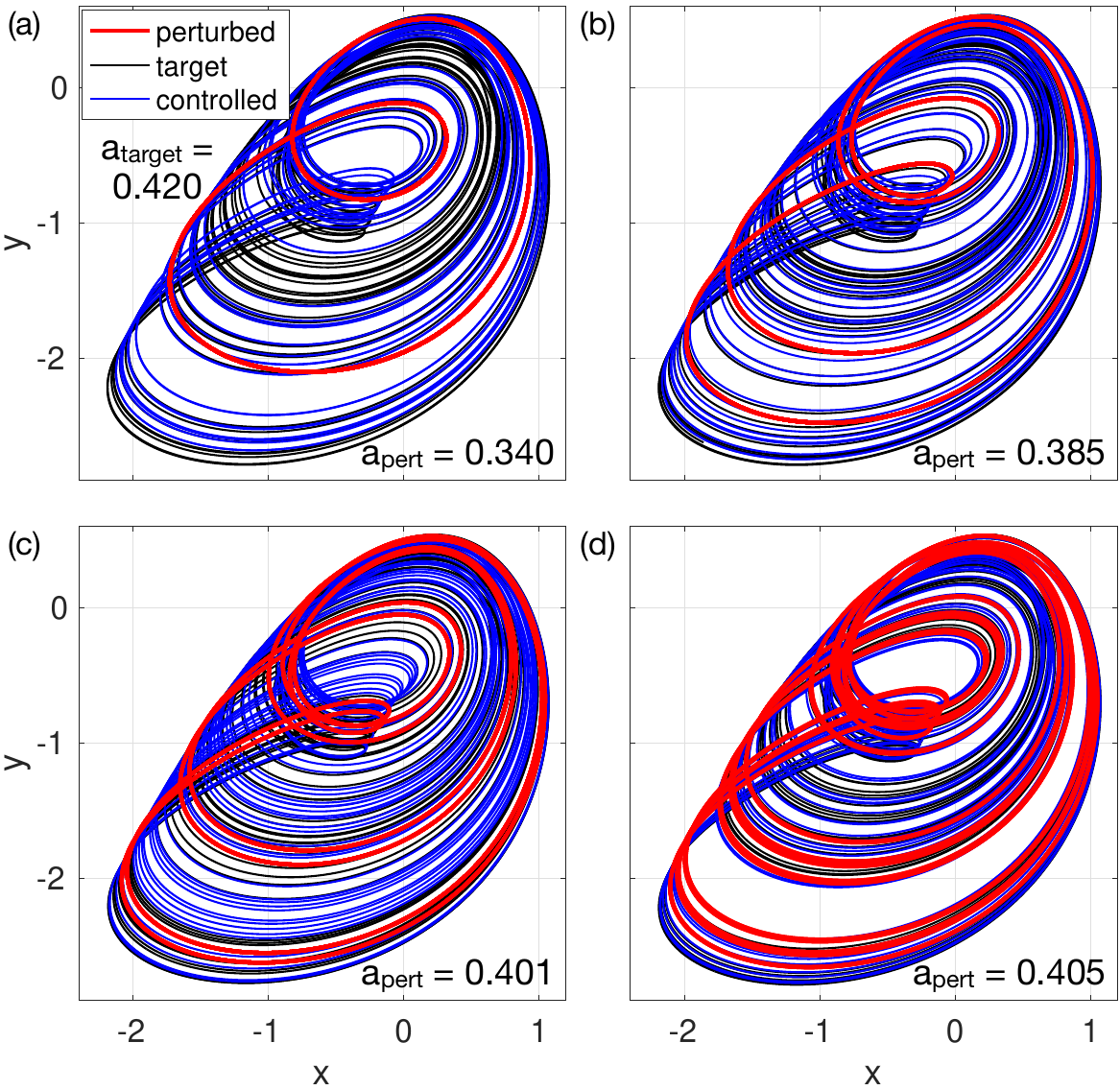}
    \caption{Attractor for the target system with $a = a_{\text{target}} = 0.420$ (black), for the system with $a = a_{\text{pert}}$ (red), and for the controlled system with $\alpha = 10$ and $\tau=4$ for $a_{\text{pert}} = 0.34$ (a), $0.385$ (b), $0.401$ (c), and $0.405$ (d) (blue).}
    \label{fig3}
\end{figure}

\begin{figure}[t]
    \centering
    \includegraphics[width = 1.0\columnwidth]{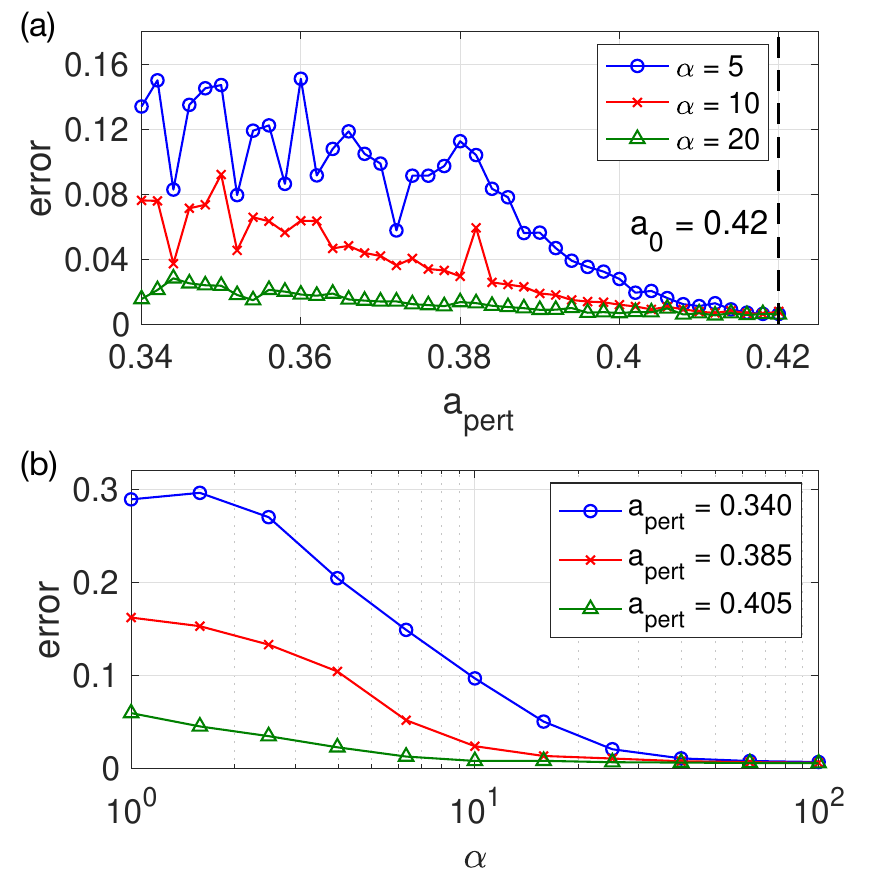}
    \caption{(a) Mean error [see Eq.~(\ref{eq:meanerror})] as a function of $a_{\text{pert}}$ for $\alpha = 5$ (blue circles), $\alpha = 10$ (red crosses), and $\alpha = 20$ (green triangles). The baseline parameter $a_0 = 0.42$ is indicated with a dashed vertical line. (b) Mean error as a function of $\alpha$ for $a_{\text{pert}} = 0.34$ (blue circles), $a_{\text{pert}} = 0.385$ (red crosses), and $a_{\text{pert}} = 0.405$ (green triangles).}
    \label{fig4}
\end{figure}

In the results shown in Fig.~\ref{fig:sprott}, the value of $a_{\text{target}}$ was such that the attractor was a limit cycle. Now we show how the dynamics can also be tuned to operate in a chaotic regime. For $a_{\text{target}} = 0.42$, there is a chaotic attractor (black curves in Fig.~\ref{fig3}). In Fig.~\ref{fig3} we show the results of tuning the dynamics to this attractor as the parameter $a$ is changed to $a_{\text{pert}} = 0.34, 0.385,0.401$, and $0.405$. These parameters would normally result in periodic orbits (red curves), but the controlled dynamics (blue curves) is chaotic. In Fig.~\ref{fig4} we again plot the mean error as a function of $a_{\text{pert}}$ and $\alpha$, noting that, again, error tends to increase as the difference $|a_{\text{pert}}-a_{\text{target}}|$ increases, but is reduced as $\alpha$ is increased. To ensure that the dynamics do not diverge, for the results in Figs.~\ref{fig3} and \ref{fig4} we reduce the amplitude of training functions in $\bm{f}$ from $0.1$ to $0.05$ and increase $\tau$ to 4. We note that preserving chaotic dynamics is fundamentally more subtle than stabilizing trajectories near periodic or equilibrium solutions, since nearby trajectories diverge exponentially and exact trajectory tracking is neither expected nor desirable.

\subsection{Parameter fluctuations in the Lorenz 63 system}

For our second example, we consider the Lorenz 63 system~\cite{lorenz1963deterministic},
\begin{align}
\frac{dx}{dt} & = \sigma(y-x),\label{lo1}\\
\frac{dy}{dt} & = x(\rho -z) -y,\label{lo2}\\
\frac{dz}{dt} & = x y - \beta z,\label{lo3}
\end{align}
where we aim to tune the dynamics to those corresponding to the baseline parameters ${\bf p}_0 = [\sigma_0, \rho_0, \beta_0]^T = [10,28,8/3]^T$. For a time step we use $\Delta t = 0.002$. In this example, we assume that all the parameters fluctuate stochastically following an Ornstein-Uhlenbeck process centered around the baseline parameters, i.e.,
\begin{align}
d \sigma = (\sigma_0 - \sigma)dt + \eta dW^{\sigma}_t,\label{pa1}\\
d \rho = (\rho_0 - \rho)dt + \eta dW^{\rho}_t,\label{pa2}\\
d \beta = (\beta_0 - \beta)dt + \eta dW^{\beta}_t,\label{pa3}
\end{align}
where $W_t^{\sigma}$, $W_t^{\rho}$, and $W_t^{\beta}$ are independent Wiener processes and we choose $\eta = 1/2$.
\begin{figure*}[t]
    \centering
    \includegraphics[width = 0.95\textwidth]{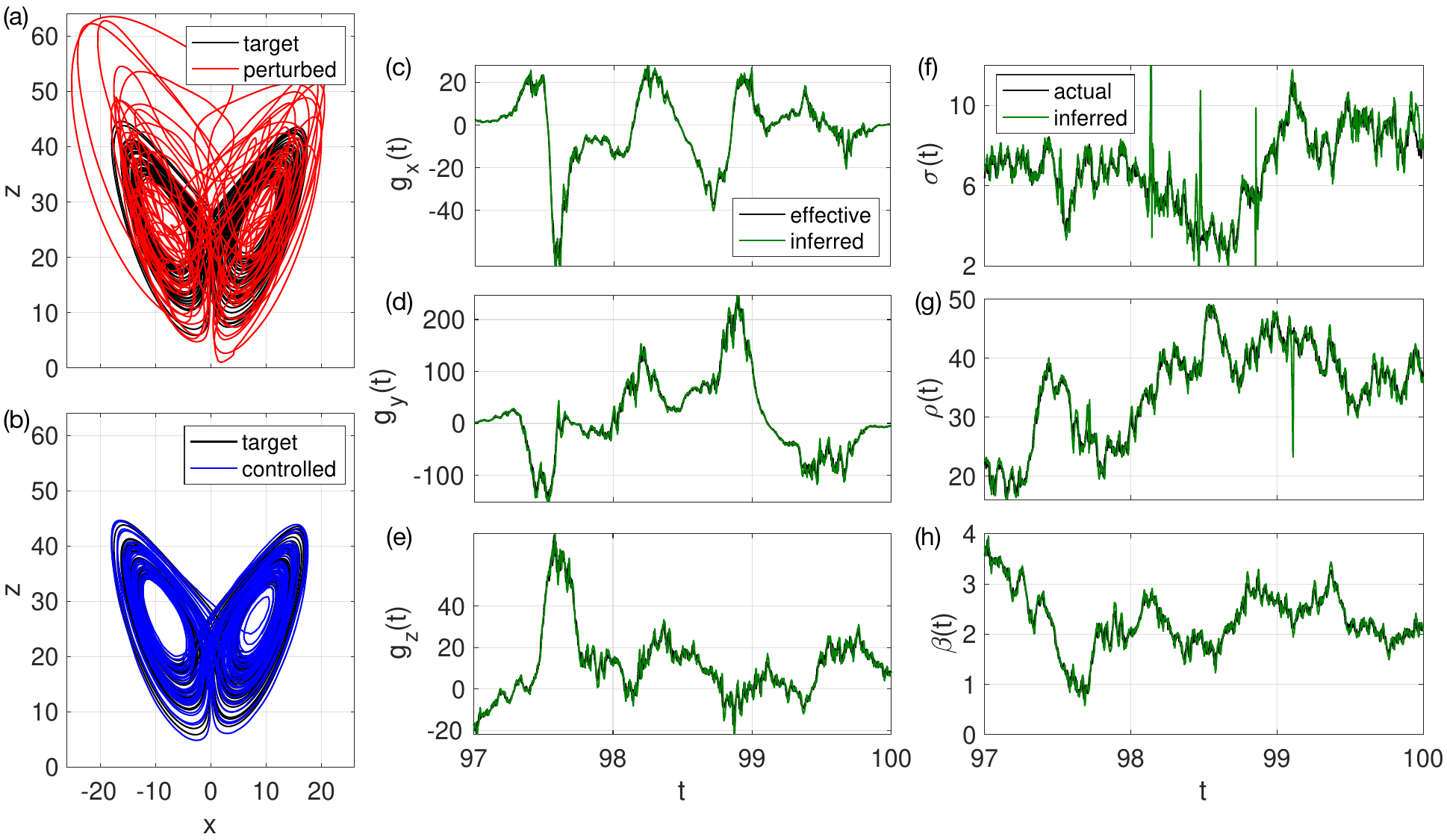}
    \caption{(a) A typical trajectory obtained from Eqs.~(\ref{lo1})-(\ref{pa3}) (red) and the Lorenz attractor obtained using the constant, baseline parameters $[\sigma,\rho,\beta] = [\sigma_0,\rho_0,\beta_0]$ (black). (b) A typical trajectory obtained by implementing Eqs.~(\ref{lop1})-(\ref{lop4}) with $\alpha = 100$ and $\tau = 1$ (blue) and the Lorenz attractor (black). (c), (d), and (e): actual perturbations $g_x$, $g_y$, and $g_z$ [see Eqs.~(\ref{gg1})-(\ref{gg3})] (black), and the identified perturbations $\alpha v_x + u_x$, $\alpha v_y + u_y$, $\alpha v_z + u_z$ using $\alpha = 100$ (green). (f), (g), and (h): actual values of the stochastically drifting parameters $\sigma(t)$, $\rho(t)$, and $\beta(t)$  (black), respectively, and the inferred values of these parameters using Eqs.~(\ref{infersigma})-(\ref{inferbeta}) (green) with $\epsilon = 0.01$.}
    \label{fig:lorenz63}
\end{figure*}

In Fig.~\ref{fig:lorenz63}(a) we show a typical disturbed trajectory obtained from Eqs.~(\ref{lo1})-(\ref{pa3}) and the Lorenz attractor obtained using the constant, baseline parameters $[\sigma,\rho,\beta] = [\sigma_0,\rho_0,\beta_0]$ in red and black, respectively. In order to tune the dynamics at the baseline parameters $[\sigma_0,\rho_0,\beta_0]$, we train the reservoir using the training function 
$${\bf f}(t) = \left[\begin{array}{c}
\text{sign}[\cos(0.05t)]\\
\text{sign}[\sin(0.05t)]\\ 
\sin(0.05t)
\end{array}
\right].
$$
We also increase our control gain to $\alpha = 100$. We then implement the control scheme
\begin{align}
\frac{dx}{dt} & = \sigma(t)(y-x) - \alpha v_x,\label{lop1}\\
\frac{dy}{dt} & = x[\rho(t) -z] -y - \alpha v_y,\label{lop2}\\
\frac{dz}{dt} & = x y - \beta(t) z -\alpha v_z\label{lop3},\\
\frac{d{\bf v}}{dt} &= \frac{1}{\tau}\left( {\bf u} - {\bf v}  \right) \label{lop4},
\end{align}
where ${\bf u}$ is the output of the reservoir.

In Fig.~\ref{fig:lorenz63}(b) we show a typical trajectory obtained by implementing Eqs.~(\ref{lop1})-(\ref{lop4}) with $\alpha = 100$, $\tau = 1$ (blue) and the Lorenz attractor. The control is effective in tuning the dynamics at the baseline parameters $[\sigma_0,\rho_0,\beta_0]$ even in the face of stochastic parameter variation.

Now we illustrate the identification of the  perturbation ${\bf g}$ in real time, while the system is controlled. From Eq.~(\ref{identify}), we have
\begin{align}
 g_x &= [\sigma(t)-\sigma_0](y-x) \approx \alpha v_x + u_x,\label{gg1}\\
g_y &= [\rho(t)-\rho_0] x \approx \alpha v_y + u_y,\label{gg2}\\
g_z &= -[\beta(t)-\beta_0]z \approx  \alpha v_z + u_z.\label{gg3}
\end{align}
Figures~\ref{fig:lorenz63}(c), (d), and (e) show the actual perturbations $g_x = [\sigma(t)-\sigma_0][y(t) - x(t)]$, $g_y = [\rho(t)-\rho_0] x(t)$, and $g_z = -[\beta(t)-\beta_0]z(t)$, as a function of $t$ and the identified disturbances $\alpha v_x + u_x$, $\alpha v_y + u_y$, $\alpha v_z + u_z$ in black and green, respectively. The reservoir succeeds in identifying the perturbation to the system. 

If we further assume that the system dynamics and baseline parameters are known, we can determine approximately the values of the drifting parameters in real time while the system is controlled.
In the notation of Eq.~(\ref{eq:G}), we need to solve  
\begin{align}
\left[
\begin{array}{ccc}
y(t)-x(t) & 0 & 0\\
0 & x(t) & 0 \\
0 & 0 & -z(t)
\end{array}
\right]
\left[
\begin{array}{c}
\sigma(t) -\sigma_0\\
\rho(t) - \rho_0 \\
\beta(t) -\beta_0
\end{array}
\right]
=
\left[
\begin{array}{c}
\alpha v_x + u_x\\
\alpha v_y + u_y\\
\alpha v_z + u_z
\end{array}
\right]
\nonumber.
\end{align}
Since the matrix on the left-hand side is non-invertible when $x(t) = 0$, $z(t) = 0$, or $x(t) - y(t) = 0$, we introduce a small regularization parameter $\epsilon$ to approximately solve for the parameters:
\begin{align}
\sigma(t) &\approx \sigma_0 + (\alpha v_x(t) + u_x(t))\frac{[y(t)-x(t)]}{[y(t)-x(t)]^2 +\epsilon},\label{infersigma}\\
\rho(t) &\approx \rho_0 +(\alpha v_y(t) + u_y(t)) \frac{x(t)}{x(t)^2 + \epsilon},\label{inferrho}\\
\beta(t) &\approx \beta_0 - (\alpha v_z(t) + u_z(t))\frac{z(t)}{z(t)^2 + \epsilon}. \label{inferbeta}
\end{align}
In Figs.~\ref{fig:lorenz63}(f), (g), and (h) we show the actual values of the stochastically drifting parameters $\sigma(t)$, $\rho(t)$, and $\beta(t)$  (black), and their inferred values, respectively, using Eqs.~(\ref{infersigma})-(\ref{inferbeta}) (green) with $\epsilon = 0.01$. Since in our simulations the variable $z(t)$ does not become zero, the reconstruction of $\beta(t)$ is very good without the need for any regularization. In contrast, the reconstruction of $\sigma(t)$ and $\rho(t)$ improves when a nonzero regularization parameter $\epsilon$ is used. 

We emphasize that the fluctuating parameters are inferred in real time and while the system is being controlled. Furthermore, the reconstruction algorithm does not rely on the numerical estimation of the time derivatives of the variables.


\subsection{Lorenz 96 system}

For our last example, we consider a system of higher dimensionality, the Lorenz-96 system~\cite{lorenz1996predictability}, consisting of $N$ differential equations for the variables $x_1, x_2,\dots,x_N$ given by
\begin{align}
\frac{dx_n}{dt} = (x_{n+1}-x_{n-2})x_{n-1} - x_n + F_n, \label{eq:L96}
\end{align}
where periodic boundary conditions $x_0 = x_N$, $x_{-1} = x_{N-1}$, $x_{N+1} = x_1$ are assumed and we use $N = 8$. To break the rotational symmetry, we allow the parameters $F_n$ to be heterogeneous, and use $F_1 = F_2 = (4/5)F$, $F_n = F$ for $n >2$. For a time step we use $\Delta t = 0.01$.

\begin{figure}[t]
    \centering
    \includegraphics[width = 1\columnwidth]{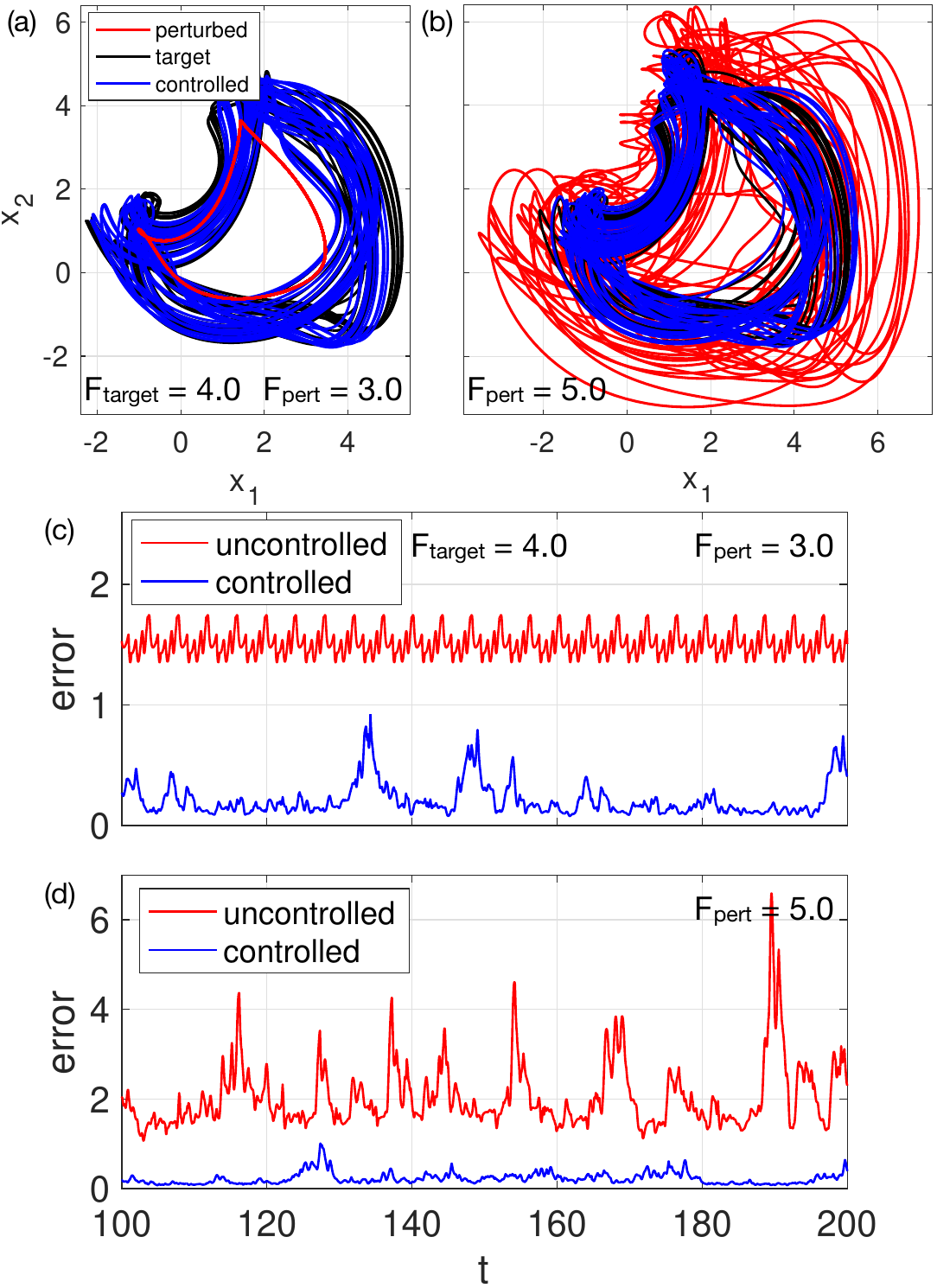}
    \caption{Panels (a) and (b) show the attractor of Eqs.~(\ref{eq:L96}) projected onto the $(x_1,x_2)$ plane for $F = F_{\text{target}} = 4.0$ (black curves). The perturbed attractor is shown in red for $F = F_{\text{pert}} = 3.0$ (a) and $F = F_{\text{pert}} = 5.0$ (b). Panels (c) and (d) show the instantaneous error [see Eq.~(\ref{eq:error})] as a function of time for the uncontrolled (red) and controlled (blue) systems, for $F_{\text{pert}} = 3.0$ (c) and  $F_{\text{pert}} = 5.0$ (d).  $\alpha = 100$.}
    \label{fig6}
\end{figure}

The black curves in Figs.~\ref{fig6}(a) and (b) show the attractor corresponding to Eqs.~(\ref{eq:L96}) projected onto the $(x_1,x_2)$ plane for $F = F_{\text{target}} = 4.0$. The attractor appears to be mildly chaotic. When $F = F_{\text{pert}} = 3.0$, the system settles at a periodic orbit [red curve in Fig.~\ref{fig6}(a)]; on the other hand, when $F = F_{\text{pert}} = 5.0$, the attractor expands substantially [red curve in Fig.~\ref{fig6}(b)]. In both cases, applying the control scheme in Eqs.~(\ref{eq1})-(\ref{eq2}) brings the system (blue curves) close to the original attractor.

Figures~\ref{fig6}(c) and (d) show the instantaneous error, defined as the distance of the trajectory to the attractor,
\begin{align}
\text{error}(t) =  \min_j\| {\bf{x}}(t) - \tilde{\bf {x}}(j \Delta t) \|,\label{eq:error}
\end{align}
where $\{\tilde{\bf x}(j \Delta t)\}$ is a time series of an orbit in the attractor with $F = F_{\text{target}}$. The error for the uncontrolled system is shown in red, and the error for the controlled system is shown in blue. In both cases, the controlled system remains much closer to the target attractor than the uncontrolled system.

\section{Discussion} \label{sec4}

In this paper, we proposed and demonstrated a model-free method for maintaining a dynamical system near a prescribed dynamical regime when the system undergoes a deterministic or stochastic disturbance in parameters. Unlike external additive disturbances applied to a fixed dynamical system, parameter variations modify the governing dynamics themselves and can induce qualitative transitions such as bifurcations and changes in attractor structure. Our results demonstrate that reservoir-computing-based disturbance identification can be extended to this setting, allowing the suppression of effective perturbations induced by parameter drift and the preservation of target dynamics.

The method requires only observations of the original system under known perturbations and does not require knowledge of the governing equations or of the parameter changes. In our implementation, a reservoir computer is trained to identify perturbations to the dynamics from observations of the system state when forced by known perturbations. Once trained, the reservoir is able to identify in real time effective perturbations arising from parameter changes, even though these perturbations are state dependent and were not explicitly included during training. We demonstrated this approach in systems exhibiting periodic and chaotic dynamics, including situations where parameter drift would otherwise induce transitions between qualitatively different dynamical regimes.

When the governing equations are known, the identified perturbation can also be used to infer the drifting parameters in real time while the system remains controlled. In the Lorenz example, the method successfully reconstructed stochastically varying parameters without requiring numerical estimation of time derivatives.

The delayed feedback control scheme used in this work was introduced previously in the context of suppressing external disturbances to dynamical systems~\cite{restrepo2024suppressing}. Here we showed that the same framework can also suppress parameter-induced perturbations, despite the fact that these perturbations arise from modifications of the intrinsic dynamics. Explaining the ability of the reservoir to generalize from simple, external training perturbations to state-dependent perturbations induced by parameter drift remains an interesting open question.

Several limitations and directions for future work remain. First, while the method successfully preserves trajectories close to the target dynamical regime, particularly in periodic systems, a more systematic characterization of the controlled chaotic dynamics would be valuable. For example, future work could compare invariant measures, Lyapunov exponents, or spectral properties between the target and controlled attractors. Second, the performance of the method is expected to depend on the observability of the perturbation from the measured variables, the choice of training perturbations, and the expressive capacity of the reservoir computer. Third, although the parameter reconstruction procedure works well in the examples considered here, the inversion problem may become ill-conditioned or non-unique in more complicated systems.

More broadly, the framework introduced here may be useful in applications where maintaining a desired dynamical regime is important despite uncertain or drifting parameters. Potential examples include power-grid dynamics, ecological systems, fluid systems, and other nonlinear systems operating near bifurcation points or under slowly varying environmental conditions.



\begin{acknowledgements}

JGR acknowledges support from {the National Science Foundation (grant number DMS-2205967).}

\end{acknowledgements}




\bibliography{refs.bib}


\end{document}